\documentclass[onecolumn,12pt,nofootinbib,showpacs,aps]{revtex4}

\usepackage{graphicx}
\usepackage{bm}
\usepackage{dsfont}
\usepackage{mathrsfs} 
\usepackage{amssymb,amsfonts,amsmath}
\usepackage{graphicx}
\usepackage{datetime}
\usepackage{color}
\usepackage{amsthm}

\allowdisplaybreaks[4]

\newcommand{\dd}{\mathrm{d}}

\newcommand{\Ae}{A_{\textrm{E}}}
\newcommand{\AC}{A_{\textrm{C}}}
\newcommand{\AEq}{\bar A_{\textrm{E}}}
\newcommand{\ACq}{\bar A_{\textrm{C}}}

\normalsize

\theoremstyle{plain}
\newtheorem{thm}{Theorem}
\begin{document}

\title{\Large Geometric relations for rotating and charged\\ AdS black holes}

\author{\bf J\"org Hennig}
\email{jhennig@maths.otago.ac.nz}

\affiliation{Department of Mathematics and Statistics,
           University of Otago,
           P.O. Box 56, Dunedin 9054, New Zealand}


\begin{abstract}
 We derive mass-independent equations and inequalities for Kerr-Newman-anti-de Sitter black holes. In particular, we obtain an equation that relates electric charge, angular momentum and the areas of the event and Cauchy horizons. An area-angular momentum-charge inequality is derived from this formula, which becomes an equality in the degenerate limit. The same equation is shown to hold for arbitrary degenerate black holes, which might, for example, be surrounded by matter.
 
\end{abstract}

\maketitle
\section{Introduction \label{sec:intro}}

For the Kerr-Newman (KN) black hole, it is easy to see that the areas $\Ae$ and $\AC$ of the event and Cauchy horizons depend on all three black hole parameters $M$ (mass), $J$ (angular momentum) and $Q$ (electric charge), whereas the area product is mass-independent,
\begin{equation}\label{eq:KN}
 \AC\Ae=(8\pi J)^2+(4\pi Q^2)^2.
\end{equation}
Since all quantities in this equation can be computed locally at the horizons, one might hope that the same equation could even hold for more general black hole spacetimes, e.g.\ black holes with surrounding matter. On the other hand, relations that depend on the black hole mass should be expected to be very specific for the particular class of solutions as there is no preferred quasi-local mass concept for black holes that could replace $M$ in more general solutions. Instead, $M$ would probably become a global quantity that necessarily requires information about the surroundings of the black hole. For that reason, mass-independent, purely local, equations like \eqref{eq:KN}, as well as mass-independent inequalities, are particularly interesting. And it has been shown  that \eqref{eq:KN} does indeed hold for general axisymmetric and stationary black holes \cite{AnsorgHennig2008,AnsorgHennig2009,AnsorgHennigCederbaum2011,HennigAnsorg2010}. 

Intimately related to the horizon equation \eqref{eq:KN} is an inequality for the KN solution,
\begin{equation}\label{eq:KNineq}
 (8\pi J)^2+(4\pi Q^2)^2\le\Ae^2.
\end{equation}
This follows immediately from \eqref{eq:KN} using that the KN-horizon areas satisfy $\AC\le\Ae$, as can easily be verified from the explicit solution. Again it turns out that this result is not limited to the very special KN black hole. First it was proved that \eqref{eq:KNineq} also holds for general axisymmetric and stationary black holes with surrounding matter \cite{HennigAnsorgCederbaum2008,HennigCederbaumAnsorg2010, AnsorgHennigCederbaum2011}. Subsequently, the assumption of stationarity was dropped and the inequality was shown for general axisymmetric black holes in dynamical spacetimes \cite{JaramilloReirisDain2011, DainJaramilloReiris2012, Dain2012, GabachJaramillo2012, GabachJaramilloReiris2013}. 
As an application of the inequality \eqref{eq:KNineq} (in the case without charge, $Q=0$) we mention that it is an essential ingredient in the proof that equilibrium configurations with two aligned and rotating black holes do not exist \cite{NeugebauerHennig2009,HennigNeugebauer2011,NeugebauerHennig2012,Chrusciel2011}, i.e.\ the spin-spin repulsion is not strong enough to compensate the omnipresent mass attraction.

The next step is an extension to spacetimes with non-vanishing cosmological constant $\Lambda$. Inequality \eqref{eq:KNineq} is even valid for $\Lambda>0$ \cite{GabachJaramillo2012}, even though it then does not provide an optimal bound anymore as $\Lambda$ does not explicitly appear. On the other hand, \eqref{eq:KNineq} does not generally hold for $\Lambda<0$. Area-charge inequalities for both positive and negative cosmological constant have been derived in \cite{Simon2012}. These are valid for non-rotating and rotating black holes, but since the angular momentum does not appear in these inequalities, they cannot be expected to be optimal bounds for rotating black holes and a refinement of the inequalities would be desirable. Besides inequalities, horizon equations similar to \eqref{eq:KN} for the non-rotating, static black holes described by the Schwarzschild-(anti-)de Sitter and Reissner-Nordstr\"om-\mbox{(anti-)}de Sitter solutions have been derived in \cite{Visser2013}. These results can also be generalized to higher dimensional static black holes, see \cite{WangXuMeng2013} and references therein. 

An interesting feature of the case $\Lambda>0$ is the appearance of cosmological horizons in addition to the black hole horizons. It turns out that the areas of these new horizons must necessarily enter the horizon equations\footnote{A mass-independent formula for Reissner-Nordstr\"om-de Sitter black holes was derived in \cite{Visser2013}, cf.\ Eq.~(64) there. This equation relates the charge of the black hole to the areas of the event, Cauchy and cosmological horizons --- and it is the unique formula that only involves these quantities but not the mass. Interestingly, the three types of areas appear symmetrically in this equation, i.e.\ the nature of the horizons is irrelevant in that context, and all of them are needed. Given this observation for a family of \emph{non-rotating} de Sitter black holes, it cannot be expected that there are mass-independent formulae in the more complicated \emph{rotating} case that do not depend on the areas of cosmological horizons.}. Hence these equations are in some sense ``less local'' than the relations without cosmological horizons, as  knowledge about possibly very distant cosmological horizons is required.

In the present paper we incorporate angular momentum into the black hole relations and our main result is a generalization of the horizon equation \eqref{eq:KN} and the inequality \eqref{eq:KNineq} to the \emph{Kerr-Newman-anti-de Sitter solution}. Thereby, we focus on a negative cosmological constant, $\Lambda<0$, in order to avoid the above mentioned complications through cosmological horizons. Instead we try to be as close as possible to the situation with $\Lambda=0$ which has exactly two horizons (the event and Cauchy horizons). This is done in Sec.~\ref{sec:KN-AdS}. Afterwards, in Sec.~\ref{sec:degen}, we consider arbitrary \emph{degenerate} black holes (which might be surrounded by matter or other black holes) and prove a universal horizon equation for them, which is obtained as a simple consequence of a difficult result due to Kunduri and Lucietti \cite{Kunduri2009}. Finally, we discuss our results in Sec.~\ref{sec:discuss}.

\section{Kerr-Newman-AdS black holes\label{sec:KN-AdS}}

The Kerr-Newmann-anti-de Sitter (KN-AdS) solution in Boyer-Lindquist-type coordinates $(r,\theta,\phi,t)$ has the line element
\begin{equation}
 \dd s^2 = \varrho^2\left(\frac{\dd r^2}{\Delta_r}+\frac{\dd\theta^2}{\Delta_\theta}\right)
 +\frac{\Delta_\theta\sin^2\!\theta}{\Xi^2\varrho^2}\left[a\,\dd t-(r^2+a^2)\,\dd\phi\right]^2
 -\frac{\Delta_r}{\Xi^2\varrho^2}\left(\dd t-a\sin^2\!\theta\,\dd\phi\right)^2,
\end{equation}
where
\begin{equation}
 \varrho^2=r^2+a^2\cos^2\!\theta,\quad
 \Xi = 1+\frac{\Lambda}{3}a^2,
\end{equation}
\begin{equation}
 \Delta_r=(r^2+a^2)\Big(1-\frac{\Lambda}{3}r^2\Big)-2mr+q^2,\quad
 \Delta_\theta=1+\frac{\Lambda}{3}a^2\cos^2\!\theta,
\end{equation}
and the corresponding vector potential for the electromagnetic field is
\begin{equation}
 A=-\frac{qr}{r^2+a^2\cos^2\!\theta}(\dd t-a\sin^2\!\theta\,\dd\phi).
\end{equation}
The solution depends on the cosmological constant $\Lambda<0$ as well as on the mass parameter $m$, the rotation parameter $a$ and the charge parameter $q$, which are related to the mass $M$, the angular momentum $J$ and the electric charge\footnote{We only consider black holes with a purely electric charge, i.e.\ without magnetic charge.} $Q$ via
\begin{equation}
 M=\frac{m}{\Xi^2},\quad 
 J=\frac{ma}{\Xi^2}\equiv Ma,\quad
 Q=\frac{q}{\Xi}.
\end{equation}
Note that $a$ must be chosen small enough so that $-\frac13\Lambda a^2<1$, which is necessary in order to guarantee a metric with Lorentzian signature for all coordinate values.

The zeros of $\Delta_r$, which is a quartic in $r$, give the coordinate radii of the horizons. For positive $\Lambda$ there are up to four real zeros, corresponding to black hole horizons and  cosmological horizons. In the present case of negative $\Lambda$, however, there are at most two real zeros $r_1$ and $r_2$, corresponding to an outer event horizon and an inner Cauchy horizon. Since we are considering black holes, we will assume that $a$ and $q$ are chosen sufficiently small to guarantee that there are indeed two zeros (for a subextremal black hole) or one double zero (corresponding to a degenerate black hole). For larger $a$ and $q$, $\Delta_r$ may have no real zeros, in which case we have a solution with a naked singularity rather than a black hole.

We also note that the surface gravity is given by
\begin{equation}\label{eq:kappa}
 \kappa=\frac{1}{2(r^2+a^2)}\frac{\dd}{\dd r}\Delta_r\big|_{r=r_2},
\end{equation}
where $r_2$ is the radial coordinate of the event horizon (the larger of the two real zeros of $\Delta_r$). This shows that a degenerate black hole, which is defined by vanishing surface gravity, $\kappa=0$, is indeed characterized by a double zero of $\Delta_r$, since $\Delta_r$ and $\frac{\dd}{\dd r}\Delta_r$ must vanish simultaneously at the event horizon in this case.

For the quartic equation $\Delta_r=0$ we can write down four Vieta relations for the four \emph{complex} zeros $r_1,\dots,r_4$, i.e.\ we compare the coefficients of different $r$-powers in this equation with the coefficients of $-\frac{\Lambda}{3}(r-r_1)(r-r_2)(r-r_3)(r-r_4)=0$. We choose $r_1<r_2$ to be the two real zeros and $r_3=\bar r_4$ the two complex conjugate zeros. The Vieta relations are
\begin{eqnarray}
  0 &=& r_1+r_2+2\Re(r_3),\\
  1 &=& \lambda [2\Re(r_3)(r_1+r_2)+r_1r_2+\Re(r_3)^2+\Im(r_3)^2-a^2],\\
  2m &=& \lambda\left[(r_1+r_2)[\Re(r_3)^2+\Im(r_3)^2]+2\Re(r_3)r_1r_2\right],\\
  a^2+q^2 &=& \lambda\left[\Re(r_3)^2+\Im(r_3)^2\right]r_1r_2,
\end{eqnarray}
where we have defined
\begin{equation}
 \lambda=-\frac{\Lambda}{3}>0.
\end{equation}
Since the complex roots have no physical meaning, we combine these four equations such that $\Re(r_3)$ and $\Im(r_3)$ are eliminated. In this way we obtain the two equations
 \begin{eqnarray}\label{eq:realEqn1}
  2m &=& (r_1+r_2)\left[1+\lambda(a^2+r_1^2+r_2^2)\right],\\
  \label{eq:realEqn2}
  a^2+q^2 &=& \left[1+\lambda(a^2+r_1^2+r_1r_2+r_2^2)\right]r_1r_2.
 \end{eqnarray}
We intend to use these equations to derive a mass-independent formula for the areas $\Ae$ and $\AC$ of event and Cauchy horizons, which can be computed from $r_1$ and $r_2$ via
\begin{equation}\label{eq:areas}
 \AC=4\pi\frac{r_1^2+a^2}{\Xi},\quad
 \Ae=4\pi\frac{r_2^2+a^2}{\Xi}.
\end{equation}
In the non-rotating case $a=0$, this could be done by first deriving mass-independent relations for the two zeros $r_1$, $r_2$ and then reformulating the result in terms of the areas $\AC$ and $\Ae$. In the rotating case, however, the transformation \eqref{eq:areas} from the zeros to the areas again involves the mass $M$ in the form of the parameter $a=J/M$, such that a mass-independent relation for the zeros would lead to a mass-dependent formula for the areas. Instead, we have to deal directly with the areas. To this end, we intend to eliminate $r_1$ and $r_2$ from \eqref{eq:realEqn1}, \eqref{eq:realEqn2} using \eqref{eq:areas}. In order to avoid square roots of areas, this is most conveniently done by noting that \eqref{eq:realEqn1} and \eqref{eq:realEqn2} imply
\begin{equation}
 r_1+r_2=\frac{2m}{1+\lambda(a^2+r_1^2+r_2^2)}\quad\textrm{and}\quad
 r_1 r_2=\frac{a^2+q^2-\lambda r_1^2r_2^2}{1+\lambda(a^2+r_1^2+r_2^2)}.
\end{equation}
If we substitute these expressions into the left-hand sides of the trivial identities $(r_1+r_2)^2-2r_1r_2=r_1^2+r_2^2$ and $(r_1r_2)^2=r_1^2r_2^2$, then we obtain two equations that contain only even powers of the radii. Now we can replace $r_1$ and $r_2$ by the ``reduced areas'' $\ACq$ and $\AEq$, defined by
 \begin{equation}\label{eq:reduced}
 \ACq:=\frac{\AC}{4\pi}=\frac{r_1^2+a^2}{\Xi},\quad
 \AEq:=\frac{\Ae}{4\pi}=\frac{r_2^2+a^2}{\Xi},
\end{equation}
and we also plug in $m = J\Xi^2/a$ and $q = Q\Xi$. This leads to
\begin{equation}
 4J^2=\Big[\ACq+\AEq+2Q^2+2\lambda\Big(\ACq^2+\ACq\AEq+\AEq^2+2J^2+(\ACq+\AEq)Q^2\Big)+\lambda^2(\ACq+\AEq)(\ACq^2+\AEq^2)\Big]a^2
\end{equation}
and
\begin{eqnarray}
 0 &=& -\Big[\ACq\AEq-Q^4+2\lambda\ACq\AEq(\ACq+\AEq+Q^2)+\lambda^2\ACq\AEq(\ACq^2+\ACq\AEq+\AEq^2)\Big]\nonumber\\
 &&
 +\Big[\ACq+\AEq+2Q^2
 +2\lambda\Big(\ACq^2+3\ACq\AEq+\AEq^2+(\ACq+\AEq)Q^2-2Q^4\Big)\nonumber\\
 &&\quad
  +\lambda^2(9\ACq^2\AEq+9\ACq\AEq^2+\ACq^3+\AEq^3+8\ACq\AEq Q^2)
 +4\lambda^3\ACq\AEq(\ACq^2+\ACq\AEq+\AEq^2)\Big]a^2\nonumber\\
 &&
 -3\lambda\Big[\ACq+\AEq+2Q^2
 +2\lambda\Big(\ACq^2+2\ACq\AEq+\AEq^2+(\ACq+\AEq)Q^2-Q^4\Big)\nonumber\\
 &&\quad
 +\lambda^2(\ACq^3+5\ACq^2\AEq+5\ACq\AEq^2+\AEq^3+4\ACq\AEq Q^2)
 +2\lambda^3\ACq\AEq(\ACq^2+\ACq\AEq+\AEq^2)\Big]a^4\nonumber\\
 &&
 +\lambda^2\Big[3(\ACq+\AEq)+6Q^2
 +2\lambda\Big(3\ACq^2+5\ACq\AEq+3\AEq^2+3(\ACq+\AEq)Q^2-2Q^4\Big)\nonumber\\
 &&\quad
 +\lambda^2(3\ACq^3+11\ACq^2\AEq+11\ACq\AEq^2+3\AEq^3+8\ACq\AEq Q^2)\nonumber\\
 &&\quad
 +4\lambda^3\ACq\AEq(\ACq^2+\ACq\AEq+\AEq^2)\Big]a^6\nonumber\\
 &&
 -\lambda^3\Big[\ACq+\AEq+2Q^2
  +\lambda\Big(2\ACq^2+3\ACq\AEq+2\AEq^2+2(\ACq+\AEq)Q^2-Q^4\Big)\nonumber\\
 &&\quad
  +\lambda^2(\ACq^3+3\ACq^2\AEq+3\ACq\AEq^2+\AEq^3+2\ACq\AEq Q^2)\nonumber\\
 &&\quad
  +\lambda^3\ACq\AEq(\ACq^2+\ACq\AEq+\AEq^2)\Big]a^8.
\end{eqnarray}
We can solve the former equation for $a^2$ and plug the result into the latter equation. In this way we obtain the remarkably simple horizon equation
\begin{equation}\label{eq:horEq}
 4J^2+Q^4 = \ACq\AEq
\Big[1+2\lambda(\ACq+\AEq+Q^2)+\lambda^2(\ACq^2+\ACq\AEq+\AEq^2)\Big].
\end{equation}
For $\lambda=0$, \eqref{eq:horEq} reduces immediately to $4J^2+Q^4=\ACq\AEq$, i.e.\ the horizon equation \eqref{eq:KN} (reformulated in terms of $\ACq$ and $\AEq$). For $\lambda>0$, the right-hand side is greater than $\ACq\AEq$ so that $J$ takes on a larger value than for a black hole with the same horizon areas and charge but without the cosmological constant. 
Note that due to the appearance of $Q$ on the right-hand side, this formula is not of the form where a function of the physical parameters (angular momentum and charge) alone is equal to a function of the horizon areas, as in the case $\lambda=0$.

In the special case without rotation, $J=0$, the horizon equation leads to the following expression for the charge,
\begin{equation}\label{eq:QEq}
 Q^2 =\sqrt{\ACq\AEq}\left[1+\lambda(\ACq+\sqrt{\ACq\AEq}+\AEq)\right].
\end{equation}
(Note that \eqref{eq:horEq} with $J=0$ is a quadratic for $Q^2$, which has two solutions. One of these two, however, turns out to contradict the equations $\Delta_r(r_1)=\Delta_r(r_2)=0$ and can therefore be ruled out.)
This formula was already derived in \cite{Visser2013}\footnote{There seems to be a typo in \cite{Visser2013}. The term $(\sqrt{\AC}+\sqrt{\Ae})^2$ in Eq.~(86) there should read $\AC+\sqrt{\AC\Ae}+\Ae$.}. Since the right-hand side of the latter equation is a monotonic function of $\ACq$ and using $\ACq\le\AEq$ (which follows from \eqref{eq:reduced} and $r_1\le r_2$), we immediately obtain an inequality between charge and event horizon area,
\begin{equation}\label{eq:Qineq}
 Q^2\le \AEq(1+3\lambda\AEq),
\end{equation}
where equality holds in the degenerate limit (where $\ACq=\AEq$). 
From the point of view of an external observer, such an inequality (which only depends on quantities defined on the event horizon) might be more interesting than an equation like \eqref{eq:QEq} that depends explicitly on the area $\ACq$ of the Cauchy horizon --- a quantity that is not measurable from outside the black hole.

Remarkably, it was shown in \cite{Simon2012} that this inequality is valid even in the much more general case of dynamical, charged black holes with surrounding matter (satisfying the dominant energy condition), which are modelled as stable MOTS, i.e.\ marginally outer trapped surfaces, 
(see Eq.~(7) in \cite{Simon2012}  and choose $g=0$ there as the genus for our spherical horizons).

Now we come back to the case with rotation, $J\neq 0$. Using again $\ACq\le\AEq$, we derive the following area-angular momentum-charge inequality for the KN-AdS solution from \eqref{eq:horEq},
\begin{equation}\label{eq:ineq}
 4J^2+Q^4 \le \AEq^2\Big[1+2\lambda(2\AEq+Q^2)+3\lambda^2\AEq^2\Big].
\end{equation}
In order to have a closer analogy to the inequality \eqref{eq:Qineq} from the case without rotation, we can reformulate \eqref{eq:ineq} as
\begin{equation}\label{eq:Qineq2}
 Q^2 \le \AEq\left[\sqrt{(1+2\lambda\AEq)^2-\frac{4J^2}{\AEq^2}}+\lambda\AEq\right],
\end{equation}
which also implies a bound on $J$ as the square root must be real. This imposes a sharper bound on the charge than \eqref{eq:Qineq} and might well turn out to be the correct refinement of this inequality  outside the KN-AdS family, i.e.\ for arbitrary black holes.

Note that the right-hand side of \eqref{eq:ineq} is larger for $\lambda\neq 0$ than for $\lambda=0$. This implies in particular that the inequality $4J^2+Q^2\le \AEq^2$, which holds for black holes without cosmological constant, can be violated by KN-AdS black holes (and there was no reason at all to assume that it should also hold for negative cosmological constant). This violation has already been observed in \cite{Booth2008}, based on numerical calculations.

Finally, we observe that inequality \eqref{eq:ineq} becomes an equality if and only if the black hole is degenerate, i.e.\ for $\ACq=\AEq$. Hence we have
\begin{equation}\label{eq:degen}
 4J^2+Q^4 = \AEq^2\Big[1+2\lambda(2\AEq+Q^2)+3\lambda^2\AEq^2\Big]
\end{equation}
for degenerate KN-AdS black holes. Interestingly, this formula turns out to be universal, i.e.\ it even holds for arbitrary degenerate black holes in axisymmetric equilibrium configurations (e.g.\ black holes with surrounding matter) as we will show in the next section. This supports the conjecture that inequality \eqref{eq:ineq} [or the reformulation \eqref{eq:Qineq2}] could also hold for arbitrary black holes, since these are expected to become equalities in the degenerate limit.

\section{General degenerate black holes\label{sec:degen}}

So far we have focused on very special black hole solutions with negative cosmological constant, namely KN-AdS black holes. Here we will consider the much more general situation of arbitrary degenerate and stationary black holes with $\Lambda<0$.  For example, one could think of a black hole surrounded by a ring of matter, or a black hole in a multi-black-hole configuration, or any other black hole in some nontrivial environment. The goal of this section is to show that the horizon equation \eqref{eq:degen} applies even in this general context.

It was the remarkable result of Kunduri and Lucietti \cite{Kunduri2009} that the near-horizon geometry of arbitrary degenerate axisymmetric and stationary black holes is always the one of the extremal KN-AdS solution. The only assumption is that a neighbourhood of the event horizon is electrovacuum, i.e.\ free of any matter, since the considerations in \cite{Kunduri2009} assume the electrovacuum Einstein-Maxwell equations. 
As a consequence, each formula for quantities that can be defined locally on the event horizon of a degenerate black hole (like $J$, $Q$ and $\AEq$) will be the same as for the extremal KN-AdS solution. This already implies that \eqref{eq:degen} must hold for arbitrary degenerate black holes. But in order to verify this more explicitly, we use that $J$, $Q$ and $\AEq$ for arbitrary degenerate black holes can always be expressed as (cf.\ Eq.~(90) in \cite{Kunduri2009})
\begin{equation}\label{eq:degpars}
 J=\frac{r_{+}(1+2\lambda r_{+}^2+\lambda a^2)a}{\Xi^2},\quad
 Q=\frac{q}{\Xi},\quad
 \AEq=\frac{r_{+}^2+a^2}{\Xi},\quad
 \Xi:=1-\lambda a^2
\end{equation}
in terms of parameters $r_+$, $a$ and $q$, which must be chosen subject to the constraint (cf.\ Eq.~(62) in \cite{Kunduri2009})
\begin{equation}\label{eq:constraint}
 a^2=\frac{r_+^2(1+3\lambda r_+^2)-q^2}{1-\lambda r_+^2}.
\end{equation}
Plugging \eqref{eq:degpars} into the horizon equation \eqref{eq:degen} and using the constraint \eqref{eq:constraint}, it follows that the equation is identically satisfied. Hence we arrive at the following.
\begin{thm}\label{thm}
 Consider a four-dimensional, axisymmetric and stationary, degenerate black hole (with an event horizon of $\mathbb S^2$ topology) in a spacetime with negative cosmological constant, \mbox{$\Lambda\equiv-3\lambda<0$}, such that a neighbourhood of the event horizon is electrovacuum. Then the angular momentum $J$, the charge $Q$ and the event horizon area $\Ae\equiv4\pi\AEq$ of the black hole satisfy the universal horizon equation
 \begin{equation}\label{eq:thm}
  4J^2+Q^4 = \AEq^2\Big[1+2\lambda(2\AEq+Q^2)+3\lambda^2\AEq^2\Big].
 \end{equation}  
\end{thm}
Since this theorem is a straightforward consequence of Kunduri and Lucietti's above mentioned equations for the black hole quantities and the constraint for the parameters, one could say that \eqref{eq:thm} is already implicitly contained in their paper \cite{Kunduri2009}. The point here was to explicitly work out this equation.

Note that in the special case without cosmological constant, $\lambda=0$, the horizon equation reduces to $4J^2+Q^4=\AEq^2$. This equation has already been shown to hold for general axisymmetric, stationary and equatorially symmetric degenerate black holes by Ansorg and Pfister \cite{AnsorgPfister2008}.

\section{Discussion \label{sec:discuss}}

We have derived a mass-independent horizon equation for Kerr-Newman-AdS (KN-AdS) black holes [Eq.~\eqref{eq:horEq}], which relates the angular momentum, the electric charge as well as the areas of the event and Cauchy horizons. This equation implies inequalities for quantities defined at the event horizon [\eqref{eq:ineq}, \eqref{eq:Qineq2}]. For vanishing cosmological constant, $\Lambda=0$, these reduce to well-known inequalities that hold not only for the Kerr-Newman solution, but for arbitrary black holes in dynamical spacetimes. Hence there might be some hope that also the inequalities shown here might turn out to be universal and not only valid for the KN-AdS black hole solutions. 

Moreover, we have derived a universal equation between angular momentum, electric charge and event horizon area for arbitrary degenerate, axisymmetric and stationary black holes (e.g.\ degenerate black holes with surrounding matter), cf.~Thm.~\ref{thm}. Since this relation is precisely the equality case of the considered inequality, the observation that this relation holds for arbitrary black holes gives further support to the conjecture that the inequality might also hold in general.

Note that the proof of the area-angular momentum-charge inequality \eqref{eq:KNineq} for general dynamical black holes (which does hold for $\Lambda\ge 0$ --- even though it is not a sharp bound for $\Lambda>0$ --- and which is generally not valid for $\Lambda<0$) is based on considerations for marginally outer trapped surfaces. The condition that these surfaces be stably outermost, together with the Einstein equations, gives rise to an integral inequality that can be studied with methods from the calculus of variations. In order to obtain the integral inequality, two manifestly nonnegative terms in the stability condition are neglected, namely a matter term (which is nonnegative due to the dominant energy condition) and a term proportional to the cosmological constant (which is nonnegative under the assumption $\Lambda\ge 0$), see \cite{GabachJaramillo2012} and references therein for details. The latter term must be included in the analysis for $\Lambda<0$, as the estimate fails if this term is negative. However, this additional term is expected to significantly complicate the calculations. Hence a very nice future result would be to identify a reformulation as a variational problem that could still be solved for that case. This would allow one to study whether inequality \eqref{eq:ineq}, presented here for KN-AdS black holes, does also hold in general.

Similarly, it would be interesting to study whether the horizon equation \eqref{eq:horEq} also holds for more general black holes. The proof of the corresponding statement in the case $\Lambda=0$, as given in \cite{AnsorgHennig2008,AnsorgHennig2009, HennigAnsorg2010}, makes essential use of soliton methods. More precisely, the proof uses the fact that the axisymmetric and stationary Einstein (--Maxwell) vacuum equations decouple into an ``essential part'' (consisting of second order equations for some of the metric potentials and the electromagnetic fields) and two first-order equations for the remaining metric potential that does not enter the essential equations. The first set of equations can be reformulated in the form of the complex Ernst equation \cite{Ernst1968,Ernst1968a, KramerNeugebauer1968}. Once the Ernst equation is solved, the remaining metric potential may immediately be obtained up to quadrature from the second set of equations. 
Remarkably, the Ernst equation belongs to the class of integrable equations, which are nonlinear equations that are equivalent to an associated \emph{linear} matrix problem, see, e.g.\ \cite{BelinskiZakharov1979, Neugebauer1979, Neugebauer1980}. This is the key feature that allowed  integration of the Einstein equations along the horizons and the symmetry axes to obtain an explicit formula for the area of the Cauchy horizon in terms of data at the event horizon, which finally led to the horizon formula.
Unfortunately, for $\Lambda\neq 0$, additional terms appear in the Einstein equations that destroy the nice decoupling, so that all equations need to be considered simultaneously. Hence, there is no analogue to the Ernst equation in this case, and it believed to be impossible to find a linear problem for these more general equations 
(at least, to the best of my knowledge, no one has yet been able to tackle this problem with soliton methods).
Hence --- if the horizon equation is valid at all in that case --- completely different methods must be used to prove it.


\begin{acknowledgments}
I would like to thank Chris Stevens for commenting on the manuscript.
\end{acknowledgments}


\end{document}